\documentclass{ws-p8-50x6-00}

\def\beq{\begin{equation}}
\def\eeq{\end{equation}}
\def\bea{\begin{eqnarray}}
\def\eea{\end{eqnarray}}
\def\({\left(}   
\def\){\right)}   
\def\eq#1{{Eq.~(\ref{#1})}}

\def\plb#1#2#3{    {\it Phys. Lett. }{\bf B#1} (#2) #3}
\def\prd#1#2#3{    {\it Phys. Rev. }{\bf D#1} (#2) #3}

\begin{document}

\title{pQCD at the Kinematical Boundary of its Applicability}
\author{Uri Maor}

\address{
HEP Department, School of Physics and Astronomy \\ 
Raymond and Beverly Sackler Faculty of Exact Science \\
Tel-Aviv University, Ramat Aviv, 69978, Israel \\
E-mail: maor@post.tau.ac.il}

\maketitle
\noindent
\abstracts{
\noindent
I examines the applicability and
possible need for a re-formulation of pQCD, as we know it, 
in the DIS limit of small $Q^2$ and $x$.
Gluon saturation, implied by
s-channel unitarity, and its possible
experimental signatures are critically assessed.
}
\section{Introduction}
The physics of DIS small $Q^2$ and small $x$ is
associated with
the search for gluon saturation signatures implied 
by s-channel unitarity \cite{SAT}.
We expect this transition to be preceded
by s-channel unitarity screening corrections 
(SC) which should be
experimentally visible even though the relevant 
amplitude has
not yet reached its unitarity (black disk) bound. 
We also know from our experience with
soft Pomeron physics that unitarity SC become appreciable 
at different energies for different channels. 
Specifically, the SC associated with soft diffractive
channels are considerably bigger than those associated 
with the elastic channel \cite{GLMprd}.\\
In the context of DIS and its pQCD interpretation,
we recall that while the global analysis of 
$F_2(x,Q^2)$ (or $\sigma_{tot}^{\gamma^*p}(W,Q^2)$)  
shows no significant deviations from DGLAP,
there are dedicated HERA investigations of 
$\frac{\partial F_2}{\partial ln Q^2}$
which may suggest a departure from
the DGLAP expectations when $Q^2$ and $x$ are 
sufficiently small. 
In my opinion, the question if these  
observations conclusively imply  
gluon saturation is not settled.
To this end I shall summarize in the following a 
detailed recent study \cite{GLMNF} of 
$\partial F_2/\partial \ln Q^2$ coupled to an analysis
of $J/\Psi$ photo and DIS production. 
\section{The s-channel unitarity bound}
$\frac{\partial F_2}{\partial ln Q^2}$ and 
the cross section for $J/\Psi$ 
photo and DIS production  
are linearily dependent, 
in the LLA of pQCD, on $xG(x,Q^2)$
and $[xG(x,Q^2)]^2$ respectively. 
Even though this simple dependence is not maintained at
higher perturbative orders, we can benefit 
from the comparison of $xG$,
which is obtained from a global
p.d.f. fit to the inclusive DIS data,  
with its s-channel unitarity bound. 
Since the 
bound is calculated within the framework of the dipole
approximation, it does not depend on our p.d.f. input.\\
In the dipole approximation we consider the time 
sequence of small $x$ DIS in the proton target 
rest frame. In this frame the virtual photon
fluctuates into a hadronic system 
($q \bar{q}$ to the lowest order) well
before the interaction with the target. 
Accordingly we have
\bea\label{A}
\sigma_{tot}(x,Q^2)\,=\,
\int_{0}^{1} dz\,\int d^2r_{\perp}
\mid\psi^{\gamma^{*}}(Q^2,r_{\perp},z)\mid^2\hat{\sigma}(r_{\perp},x,Q^2),
\eea
where $\psi^{\gamma^{*}}$ is the well known QED wave 
function of the $q \bar{q}$ system within the photon and  
$\hat{\sigma} = 4\pi\, Im a_{el}$ 
is the unknown hadronic total cross section of the 
$q \bar{q}$ system with
the target. $a_{el}\,\leq\,1$ is the elastic amplitude of a 
dipole, with a
size $r_{\perp} = 2/Q$ and energy $W^2 = \frac{Q^2}{x}$, 
scattering off the proton target.
The s-channel black unitarity bound on $\hat{\sigma}$ 
is obtained by fixing $a_{el}$ to equal its bound of unity.
In a similar procedure we can also calculate the unitarity 
bound for $xG$ where $q \bar{q}$ is replaced by $gg$.\\
The results of this comparison are shown in Fig.1, for
fixed $x\,=\,10^{-5}$,
compared with $xG$ as obtained from the commonly used p.d.f.
parameterizations. We observe that $xG$ obtained from GRV'94
significantly exceeds the unitarity bound for $Q^2<5 GeV^2$. 
With GRV'98 we have a slight excess over the unitarity bound, 
whereas both CTEQ5 and MRS'99 are well below the bound. 
Similar results are obtained when we
compare $\frac{\partial F_2}{\partial ln Q^2}$ 
with the corresponding unitarity bound.\\ 
In the following I shall examine the data on the logarithmic 
$Q^2$ slope of $F_2$ against NLO DGLAP without and with SC. 
The SC were calculated in the eiknal
formalism for which a LLA of the dipole 
approximation was utilized to calculate the input opacity.
In this procedure the SC diminish once $Q^2$ and/or $x$ are 
large enough. As we did not succeed to obtain a consistently
reasonable reproduction of the analyzed data with MRS'99, this
parameterization will be omitted in the following discussion.
\begin{figure}[t]
\epsfxsize=15pc          
\centerline{
\epsfbox{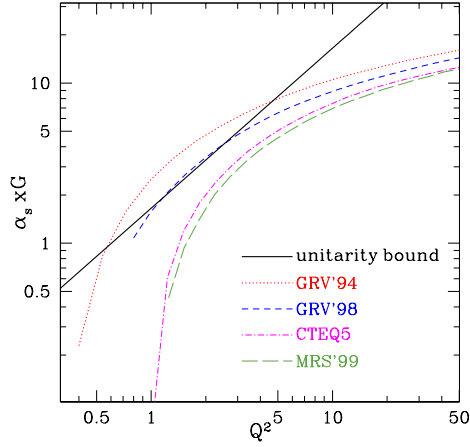}  
}
\caption{$xG$ unitarity bound at $x\,=\,10^{-5}$ compared
with various p.d.f. predictions.
\label{Fig.1}
}
\end{figure}
\section{The small $Q^2$ and $x$ behavior of 
$\frac{\partial F_2}{\partial \ln Q^2}$}
Detailed HERA data on 
$\frac{\partial F_2}{\partial ln Q^2}$
have recently became available \cite{data}. 
As I shall show, the corresponding pQCD calculation 
reproduces the data well with either 
CTEQ5HQ with no SC or with GRV'98($\overline{MS})$
corrected for SC.\\
SC were calculated in both the quark sector, to account for 
the percolation of a $q \bar{q}$ through the target, 
and the gluon sector, to
account for the screening of $xG(x,Q^2)$. 
The factorizeable result obtained is
\beq\label{B}
\frac{\partial F_2^{SC}(x,Q^2)}{\partial ln Q^2}\,=
\,D_q(x,Q^2)D_g(x,Q^2)
\frac{\partial F_2^{DGLAP}(x,Q^2)}{\partial ln Q^2},
\eeq
where the SC damping factors have been calculated
from the opacities:
$\kappa_q\,=\,\frac{2\pi \alpha_S}{3Q^2}xG^{DGLAP}(x,Q^2)\Gamma(b^2)$ 
and
$\kappa_g\,=\,\frac{4}{9}\kappa_q$.
The calculation is significantly simplified if we assume 
a Gaussian  parameterization for the two gluon non 
perturbative form factor,
$\Gamma (b^2)\,=\,\frac{1}{R^2}e^{-b^2/R^2}$, where
$R^2 = 8-10 GeV^{-2}$ is determined directly from the experimental 
forward slope of $J/\Psi$ photoproduction. 
Our results are presented
in Fig.2. We conclude that CTEQ5NSC and GRV'98SC provide 
a very good reproduction of the data and are almost identical. 
\begin{figure}[t]
\epsfxsize=18pc          
\centerline{
\epsfbox{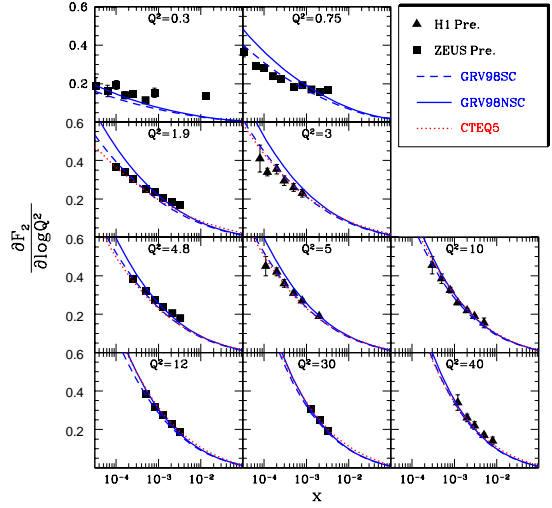} 
}
\caption{
$x$ dependence of $\partial F_2/\partial \ln{Q^2}$  
data at fixed $Q^2$.
\label{Fig.2}
}
\end{figure}
\section{Photo and DIS production of $J/\Psi$}
We try to corroborate our conclusions by examining the 
exclusive channel of $J/\Psi$ photo and DIS production.
Note that the hardness of the $J/\Psi$ channel is comparable 
to the hardness investigated in our $F_2$ study.  
The $J/\Psi$ t=0 differential cross section is
calculated in the dipole LLA. 
When using the static non relativistic approximation of 
the vector meson wave function, the cross
section has a very simple form 
\beq\label{Q}
\(\frac{d\sigma(\gamma^* p \rightarrow V p)}{dt}\)_0\,=\,
\frac{\pi^3\Gamma_{ee}M_V^3}{48\alpha}\,
\frac{\alpha_S^2{(\bar Q}^2)}{{\bar Q}^8}
\(xG(x,{\bar Q}^2)\)^2 \(1\,+\,\frac{Q^2}{M_V^2}\),
\eeq
where in the non relativistic limit we have
${\bar Q}^2\,=\,\frac{M_V^2\,+\,Q^2}{4}$
and
$x\,=\,\frac{4{\bar Q}^2}{W^2}$.\\
The integrated cross section data available for this 
channel span a relatively wide energy range \cite{data}.
To relate the integrated cross section to \eq{Q} we need to 
know $B$, the $J/\Psi$ forward differential cross section slope. 
For this we may use
the experimental values for which  
$B_{J/\Psi} \simeq \frac{R^2}{2}$.\\
The main problem with the theoretical analysis of $J/\Psi$ 
is the realization that the simple pQCD calculation needs to be corrected
for the following reasons:\\
1) The contribution of the real part of the 
production amplitude denoted.\\
2) The contribution of the skewed (off
diagonal) gluon distributions.\\
3) Relativistic effects
produced by the Fermi motion of the bound quarks. 
In our calculation we have 
considered this correction as a free parameter.\\
Our calculation of the SC for $J/\Psi$ photo and DIS production is very 
similar to the $Q^2$ logarithmic slope
calculation presented earlier.\\
Our calculations compared with the experimental 
photoproduction data are
presented in Fig.3. GRV'98SC provides 
an excellent reproduction of the data. CTEQ5NSC gives a 
lesser quality reproduction,
in particular at the lower energies, but it 
can not be ruled out. 
\begin{figure}[t]
\epsfxsize=18pc          
\centerline{
\epsfbox{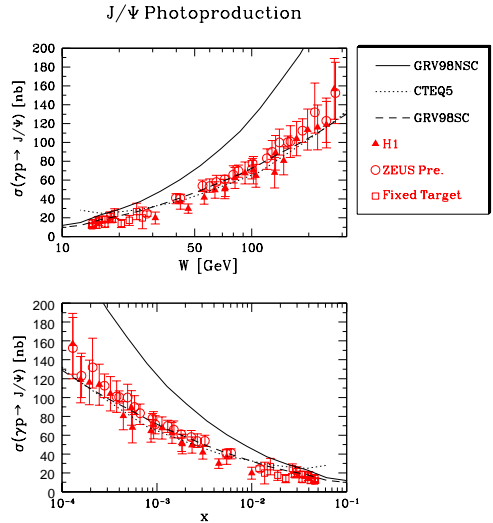} 
}
\caption{
Exclusive photo production of $J/\Psi$. 
\label{Fig.3}
}
\end{figure}
\\
Thus far we were unable to detect conclusive  
signatures of gluon saturation manifested by SC. 
Very recent HERA data \cite{data} on 
the forward x dependence of $B_{J/\Psi}$ may provide
this clue. The hard Pomeron manifested by either DGLAP 
or the dipole model is not expected to yield any shrinkage 
of the $J/\Psi$ forward photoproduction slope. SC can reproduce 
this phenomena. The two gluon form factor we have assumed,
which corresponds to  
$F_{exp} = e^{\frac{1}{4}R^{2}t}$,
provides a very mild shrinkage. However an electromagnetic 
form factor, $F_{dipole} = \frac{1}{(1\,-\,\frac{1}{8}R^2t)^2}$,
provides an excellent reproduction of the data while 
maintaining the quality of our previous results.
This is shown in Fig.4.
\begin{figure}[t]
\epsfxsize=18pc          
\centerline{
\epsfbox{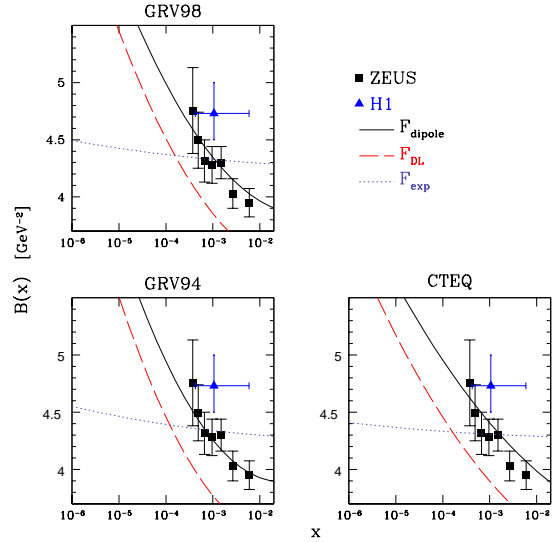} 
}
\caption{
$J/\Psi$ photoproduction forward slope.
\label{Fig.4}
}
\end{figure}
\section{Conclusions}
At this stage I do not consider the analysis presented in this 
review to provide conclusive support for the observation of gluon
saturation, even though the $B_{J/\Psi}$ data is promising.

\section*{Acknowledgments}
I wish to thank my long standing collaborators (EG,EF,EL,EN) for
their assistance.
This research was supported by in part by BSF grant \# 98000276,
by GIF grant \# I-620-22.14/1999 and by the Israel Science foundation.


\begin{thebibliography}{99}
\bibitem{data}
References to the data are given in the previous reference.
\end{thebibliography}
\end{document}